# Managing Separation of Concerns in Grid Applications Through Architectural Model Transformations

David Manset[1, 2, 3], Hervé Verjus[1], Richard McClatchey[2]

[1] University of Savoie – Polytech' Savoie – LISTIC/LS
{david.manset, herve.verjus}@univ-savoie.fr
[2] CCCS, University West of England, Bristol, UK
richard.mcclatchey@uwe.ac.uk
[3] Maat Gknowledge, Toledo, Spain
dmanset@maat-g.com

**Keywords.** Grid, MDE, Software Architecture, Model Transformation.

## 1. Introduction

Grids enable the aggregation, virtualization and sharing of massive heterogeneous and geographically dispersed resources, using files, applications and storage devices, to solve computation and data intensive problems, across institutions and countries via temporary collaborations called virtual organizations (VO) as described in [1]. Most implementations result in complex superposition of software layers, often delivering low quality of service and quality of applications. As a consequence, Grid-based applications design and development is increasingly complex, and the use of most classical engineering practices is unsuccessful. Not only is the development of such applications a time-consuming, error prone and expensive task, but also the resulting applications are often hard-coded for specific Grid configurations, platforms and infrastructures. Having neither guidelines nor rules in the design of a Grid-based application is a paradox since there are many existing architectural approaches for distributed computing, which could ease and promote rigorous engineering methods based on the re-use of software components. It is our belief that ad-hoc and semi-formal engineering approaches, in current use, are insufficient to tackle tomorrow's Grid developments requirements. Because Grid-based applications address multi-disciplinary and complex domains (health, military, scientific computation), their engineering requires rigor and control. This paper therefore advocates a formal model-driven engineering process and corresponding design framework and tools for building the next generation of Grids. To achieve these objectives, two approaches are combined: (1) a formal semantic is used to model and check Grid applications; (2) a model-driven approach is adopted to promote model re-use, through separation of concerns, to model transformations, to hide the platform complexity and to refine abstract software descriptions into concrete usable ones.

Section 2 of this paper introduces our proposal so-called *gMDE*, as well as its foundations in sections 3 and 4. Finally, section 4 and 5 illustrate the presented paradigms with an example.

## 2. A Formal Architecture-Centric MDE Approach

The presented approach, of so-called "grid Model-Driven Engineering" (*gMDE*), aims at enacting the model-driven paradigm based on formally defined architectural models dedicated to grid-based application development. While most existing MDE implementations provide only model to source code transformations where the *PIMs* are translated to *PSMs*, the problem of Grids engineering requires more elaborated models transformations – i.e. model to model - to fill the conceptual gap between the abstract model and its concrete (more detailed) representation. Moreover, interesting modelling aspects such as model optimization require the generation of intermediate models to compute and synchronize different views of the system. Thus, the proposed approach is based on the combination of the MDE vision [2] with the architecture-centric approach [3]..In Grid engineering, design is largely affected by constraints, which are introduced either by the targeted Quality of Services (QoS) or by the targeted execution platform. As presented in [4], our *gMDE* approach exhibits several models (see the right part of the Figure 1). Each model represents an accurate aspect of the system, useful for conceptual understanding (separation of concerns), analysis and refinement. Unlike the software engineering process, where the system architecture is iteratively refined by the architect, most of the transformations in *gMDE* are semi-automated. Thus, Grid applications architects only concentrate on applications functional building blocks and their interactions, and let the system address non-functional issues such as QoS. Figure 1 below, introduces the cascade of architectural models in the *gMDE* engineering process. In the presented process, a distinction is made between two major levels:

- Transformations of architectural models, which take place at the same level of abstraction (i.e. architectural structure, behaviour and properties) shown above the broken line in figure 1 and

- Transformations of abstract models to more concrete ones (including deployment in an infrastructure) shown below the broken line in Figure 1.

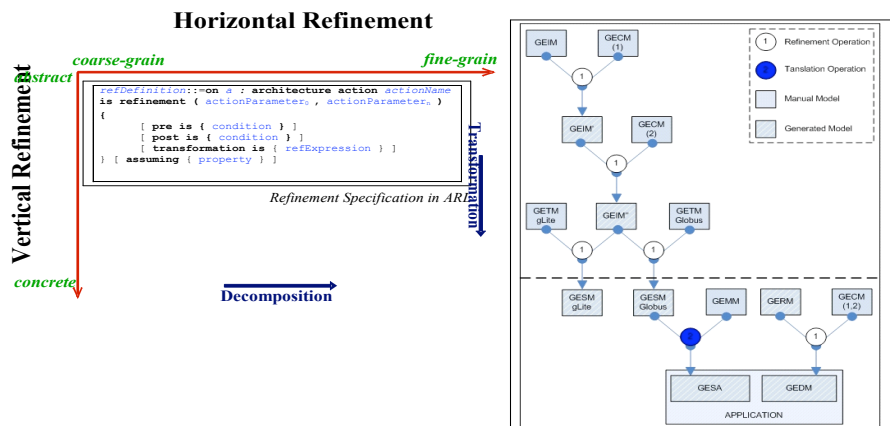

**Figure 1.** The *gMDE* Design Process as a Cascade of Refined Models

## 3. *gMDE* as a Grid-Based Application Development Framework

ArchWare [3] is an architecture-centric engineering environment supporting the development of complex systems. It enables the support of critical correctness requirements and provides languages for expressing architecture structure, behaviour and properties. ArchWare provides a set of formal languages amongst which the Architecture Refinement Language (ARL [5]). This latter is used to describe software architectures (based on the Component and Connector paradigm) and to refine them according to transformation rules. This language is based on the π-calculus [6] and µ-calculus [7] allowing the specification of architectures structure, behaviour and properties. Our *gMDE* approach uses ARL as the basis language (a refinement calculus) for expressing architectural models and transformation rules.

The *gMDE* approach focuses on both directions of refinement i.e. "vertical" and "horizontal". The intention is not only to refine an architecture to a concrete and "close to final" code form, but also to adapt it according to constraints. *gMDE* proposes two ways of using the model-driven process. The first consists of optimizing a given Grid-based application abstract architecture according to expressed developers' QoS. The second consists of adapting an architecture according to the target Grid middleware. Respectively:

- QoS. Each QoS is represented by an architectural model (considered as an architectural pattern, which can be re-used in other Grid-based application architectures). This QoS representation is then incorporated into the current Grid-based application architecture through a set of refinement actions.
- Target Grid platform. Each Grid platform is represented by another architectural pattern. The Grid-based application abstract architecture is adapted to the platform representation through a set of refinement actions too.

To address the specificities of Grids, the ARL expressiveness has to be extended: the *gMDE* approach features a Domain Specific Language (DSL) allowing the description of proper Grid services and their associated constraints. This language is based on a Grid SOA paradigm promoting simplicity and facilitating the model comprehension, architectures being naturally expressed in terms of services and their properties.

## 4. *gMDE* Architectural Models Transformation Principles

Using *gMDE*, an architect formalizes on one hand the architecture of the grid-based application (model A) and on another hand, a QoS attribute (model B). The first model is expressed by using our DSL built on top of ARL.

To engage in the weaving process, the constraint definition model (model B) is transformed into refinement actions, as illustrated in the left part of Figure 2. During the weaving process, the Grid-based application architectural model (model A) is translated in ARL and the model B is interpreted and decomposed into a series of refinement actions in ARL too. The refinement actions are applied one by one on the model A until completion, resulting in a model C satisfying the specified QoS (right part of Figure 2).

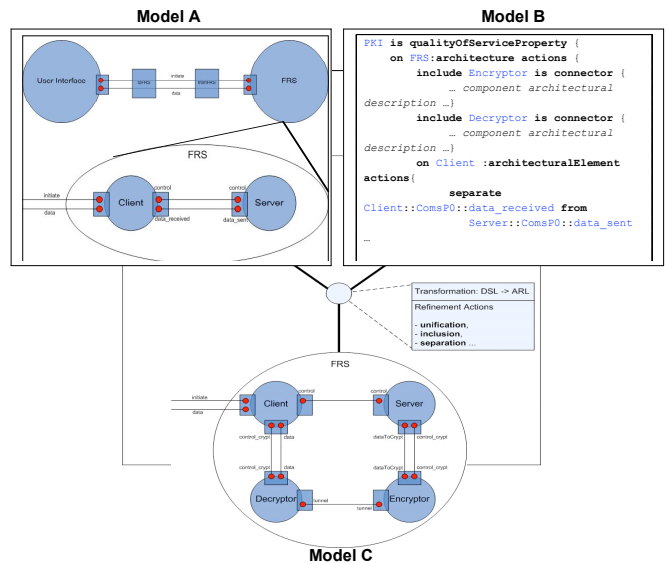

**Figure 2.** *gMDE* architectural models transformation

## 5. Conclusion

The *gMDE* approach and environment are currently in use to evaluate potential advantages in the development process of various Grid applications. There are clearly identified challenges in the development of systems such as MammoGrid [8], which can be addressed by using *gMDE*. From these case studies, preliminary conclusions are encouraging and highlight the approach relevance.

## References

[1] I. Foster, C. Kesselman, S. Tuecke., The Anatomy of the Grid – Enabling Scalable Virtual Organisations, International Journal of Supercomputer Applications, 2001.
[2] D. C. Schmidt, "Guest Editor's Introduction: Model-Driven Engineering," Computer ,vol. 39, no. 2, pp. 25-31, February, 2006.
[3] Archware, The EU funded ArchWare IST 2001-32360 – Architecting Evolvable Software - project : http://www.arch-ware.org.
[4] Manset D., Verjus H., McClatchey R., Oquendo F., « A Formal Architecture-Centric Model-Driven Approach For The Automatic Generation Of Grid Applications », Actes de 8th International Conference on Enterprise Information Systems ICEIS'06, Paphos, Chyprus, 2006.
[5] F. Oquendo, "π-ARL: an Architecture Refinement Language for Formally Modelling the Stepwise Refinement of Software Architectures", ACM Press, ACM SIGSOFT Software Engineering Notes archive Volume 29, Issue 5, September 2004.
[6] R. Milner, "Communicating and Mobile Systems: the pi-calculus", ISBN 052164320, Cambridge University Press, 1999
[7] D. Kozen, "Results on the Propositional Mu-Calculus", Theoretical Computer Science 27:333-354, 1983.
[8] S.R Amendolia et al., "Deployment of a Grid-based Medical Imaging Application", Proceedings of the 2005 HealthGrid Conference. UK, 2005.